# Boosting algorithms in energy research: A systematic review

Hristos Tyralis[1,2], and Georgia Papacharalampous[3]

[1]Department of Water Resources and Environmental Engineering, School of Civil Engineering, National Technical University of Athens, Iroon Polytechniou 5, 157 80 Zografou, Greece

[2]Air Force Support Command, Hellenic Air Force, Elefsina Air Base, 192 00 Elefsina, Greece (https://orcid.org/0000-0002-8932-4997)

[3]Department of Water Resources and Environmental Engineering, School of Civil Engineering, National Technical University of Athens, Iroon Polytechniou 5, 157 80 Zografou, Greece (https://orcid.org/0000-0001-5446-954X), (papacharalampous.georgia@gmail.com, gpapacharalampous@hydro.ntua.gr)

Corresponding author: Hristos Tyralis (hristos@itia.ntua.gr, montchrister@gmail.com)

**Abstract**: Machine learning algorithms have been extensively exploited in energy research, due to their flexibility, automation and ability to handle big data. Among the most prominent machine learning algorithms are the boosting ones, which are known to be "*garnering wisdom from a council of fools*", thereby transforming weak learners to strong learners. Boosting algorithms are characterized by both high flexibility and high interpretability. The latter property is the result of recent developments by the statistical community. In this work, we provide understanding on the properties of boosting algorithms to facilitate a better exploitation of their strengths in energy research. In this respect, (a) we summarize recent advances on boosting algorithms, (b) we review relevant applications in energy research with those focusing on renewable energy (in particular those focusing on wind energy and solar energy) consisting a significant portion of the total ones, and (c) we describe how boosting algorithms are implemented and how their use is related to their properties. We show that boosting has been underexploited so far, while great advances in the energy field are possible both in terms of explanation and interpretation, and in terms of predictive performance.

## 1. Introduction

Applications of machine learning algorithms to energy-related problems show an exponential increase, due to the rapid developments in the field of machine learning and the growing availability of big datasets [21]. Machine learning applications become more and more common in wind energy [30,40,60,90], solar energy [87,88,100], water resources [97], energy in buildings [79,85,111], renewable energy in general [81,101], fossil fuels [69], energy demand [99] and development of materials [23] to name a few disciplines.

The utility of machine learning algorithms in solving practical problems, including energy-related ones, has been highlighted in numerous studies. For instance, they are flexible, they have high predictive power, they can be fully automated (and, therefore, they can be used in operational settings), and they have been extensively tested exhibiting good performances in real world cases. Several open source implementations are available for these algorithms [73], while their combination with physics-based models [73,96] is also possible. They have been extensively benchmarked in various settings [41,72] and their properties are largely known. Although their primary focus is to generalize in previously unseen data after proper training [12], they can also be interpretable. Their strengths have been extensively exploited in energy research [21].

Boosting (an ensemble learning algorithm) holds a dominant position among the machine learning algorithms and has been described as the "*best-of-the-shelf classifier in the world*" [39 p.340]. This latter description was specifically referred to AdaBoost (abbreviation for adaptive boosting, with trees algorithm), a boosting variant that does not need extensive preprocessing and hyperparameter tuning to perform well [39 p.352,11]. The concept behind boosting originated in [82]. In this latter work, it was shown that a weak learner (i.e. a learner that performs slightly better than random guessing) can be transformed to (or "*boosted*" to become) a strong learner (worded as the "*strength of weak learnability*" at that time). This concept was later reworded as "*garnering wisdom from a council of fools*" [83 p.1]. AdaBoost [31,32,33,34] was the most successful from the early boosting algorithms.

Early boosting algorithms focused on classification, while their procedures included the sequential training of a weak learner by using reweighted versions of the training data. The prediction would be a weighted majority vote of the sequence of the trained learners. Understanding the properties of boosting procedures was mostly a focus of the machine learning community. A statistical view of boosting was demonstrated in [37]. This latter work showed that AdaBoost could be seen as an additive logistic regression model. In subsequent developments, boosting was proposed as the estimation of a function by optimizing a loss criterion via steepest gradient descent in function space [35], albeit [66] question this view. The latter developments allowed the delivery of interpretable procedures in supervised learning, and extended boosting to the regression case using arbitrary loss functions. On the other hand, some recent boosting variants (e.g. extreme gradient boosting [20], light gradient boosting [50], and categorical boosting [74]) focused on increased predictive performance and speed, thereby achieving remarkable results in real case applications and forecasting competitions. The history and the evolution of boosting algorithms can be found in [16,19,37,64,77], while reviews on the developments of boosting algorithms can be found in [9,18,65].

Considering the recent trends in energy research, manifested into both the growing number of the related machine learning applications and the success of boosting (due to its diverse properties) in several fields, in this work we aim to:

(a) Increase understanding of boosting algorithms for energy researchers.

(b) Provide a detailed review of boosting applications in energy research.

(c) Review how particular strengths of the boosting algorithms were exploited in energy studies.

To this end, we present a brief theory of the boosting algorithms, as well as their strengths and weaknesses. We systematically review approximately 150 relevant papers published in Journals specialized in energy research. We conclude with take-home remarks regarding potential applications of boosting in the energy sector.

## 2. Theory of boosting algorithms

Here we present a concise theory of boosting, while detailed descriptions can be found in books focusing on machine or statistical learning [27,39,45] or in books specialized in ensemble learning [113] in general, and in boosting in particular [83]. An overview of ensemble learning algorithms (among which boosting is a special case) can also be found in [80], while tutorials and introductory material for boosting can be found in [62,63,67].

Given the extensive use of machine learning algorithms in energy research, in what follows we assume that the reader is familiar with frequently used concepts appearing in the machine learning literature. Such concepts are the supervised and unsupervised learning, regression and classification, and training and testing, among others.

Let $\boldsymbol{x} := (x_1, \ldots, x_p)$ be the vector of predictor variables and $y$ be the dependent variable. Let also $\boldsymbol{x}_i$ be the $i^{\text{th}}$ observation, $i = 1, \ldots, n$ of $\boldsymbol{x}$ and $y_i$ be the respective $i^{\text{th}}$ observation of $y$. The focus is to estimate a function $f$: $\boldsymbol{x} \rightarrow y$, which approximates the unknown functional dependence between $\boldsymbol{x}$ and $y$ in a supervised setting.

### 2.1 Adaptive Boosting

We begin by describing AdaBoost [31,32,33,34], the most popular boosting algorithm [39 p.337]. AdaBoost is a classification algorithm using stumps (a two terminal-node decision tree) or decision trees as base learners (weak learners), albeit other options are also available [39 p.339,340]. In the two-class problem, where $y \in \{-1, 1\}$, the AdaBoost method is formulated according to Algorithm 1.

Algorithm 1. Formulation of the AdaBoost algorithm, adapted from [39 p.339].

Step 1: Initialize observation weights $w_i = 1/n$, $i = 1, \ldots, n$.

Step 2: For $m = 1$ to $M$:
a. Fit a classifier $C_m(\boldsymbol{x})$ to the training data by using weights $w_i$.
b. Compute $\text{error}_m = \sum_{i=1}^{n} w_i \, I(y_i \neq C_m(\boldsymbol{x}_i)) / \sum_{i=1}^{n} w_i$.
c. Compute $a_m = \log((1 - \text{error}_m)/\text{error}_m)$.
d. Set $w_i$ equal to $w_i \exp(a_m \, I(y_i \neq C_m(\boldsymbol{x}_i)))$, $i = 1, \ldots, n$.

Step 3: Predict $C(\boldsymbol{x}) = \text{sign}(\sum_{m=1}^{M} a_m \, C_m(\boldsymbol{x}))$ (i.e. by majority voting), where sign denotes the sign function.

The main idea behind AdaBoost is to iteratively train a weak learner (in this case $C_m(\cdot)$) to reweighted versions of the data. In each iteration, larger weights are assigned to those data that were misclassified in the previous iteration (Algorithm 1, Step 2d). This reweighting aims at forcing the algorithm to correctly classify observations that are hard to predict. Therefore, AdaBoost focuses on manipulating the data rather than the weak learner. This procedure transforms a weak classifier to a strong one.

### 2.2 Gradient boosting

AdaBoost and other early boosting algorithms were developed by the machine learning community with a focus on predictive performance. The value of boosting for statistical inference was recognized by [37], while [35] derived the gradient-descent-based formulation of boosting methods (see Algorithm 2 for this formulation [15]). This formulation is directly linked to the statistical view of boosting.

Algorithm 2. Formulation of the gradient boosting algorithm, adapted from [15,35,67].

| |
|---|
| Step 1: Initialize $f_0$ with a constant. |
| Step 2: For $m = 1$ to $M$:<br>a. Compute the negative gradient $g_m(\boldsymbol{x}_i)$ of the loss function $L$ at $f_{m-1}(\boldsymbol{x}_i)$, $i = 1, \ldots, n$.<br>b. Fit a new base learner function $h_m(\boldsymbol{x})$ to $\{(\boldsymbol{x}_i, g_m(\boldsymbol{x}_i))\}$, $i = 1, \ldots, n$.<br>c. Update the function estimate $f_m \leftarrow f_{m-1} + \rho\, h_m(\boldsymbol{x})$. |
| Step 3: Predict $f_M(\boldsymbol{x})$. |

Here one has to first decide which metric $L(y, f(\boldsymbol{x}))$ should be optimized for estimating the function $f$ (see Algorithm 2, step 2a). In the boosting terminology, this metric is usually called "loss function". Examples of loss functions can be found in Section 2.4. Second, one has to decide which base learner $h$ to use (see Algorithm 2, step 2b). Examples of base learners can be found in Section 2.5. The selection of loss functions and base learners largely depends on the research question.

Boosting works by adding in each iteration, the newly fitted base learner $h(\boldsymbol{x}, \theta_m)$ (see Algorithm 2, step 2c) to the previously estimated $f_{m-1}$. The base learner $h(\boldsymbol{x}, \theta_m)$ has been fitted in step 2b (see Algorithm 2) to the negative gradient. The stopping iteration parameter $M$ is important, since few iterations lead to unsatisfactory training (termed as "underfitting"), while too many iterations may result in overfitting (i.e. nice fitting to the training data but high probability of failing to provide good predictions of the test data). The step-length factor $\rho$ ($0 < \rho \leq 1$) is less important, while small $\rho$ values (e.g. $\rho = 0.1$) are usually satisfactory [15].

Some important remarks should be made on Algorithm 2. These are the following:

(a) The concept of boosting is to train new base learners, aiming at correlating them with the negative gradient of the loss function [67]. To get an idea, the negative gradient of the loss function $L_2$ is the error function; therefore, minimization of the loss function $L_2$ is achieved by sequentially training the base learners to the errors of the model.

(b) The term "stagewise additive expansion", used by [35], does not imply that the model is additive in the predictor variables. Instead, it implies that the model is additive with respect to the sequentially trained functions [15].

(c) Regarding the estimation of the optimal value of $M$, the most common approaches are based either on an early stopping rule within a cross validation setting (by optimizing the empirical risk, i.e. the loss function averaged over the observations) or on the use of model selection criteria (e.g., the Akaike Information Criterion, Bayesian Information Criterion) [18]. On the other hand, [104] suggest that the boosting algorithms should be used without early stopping, and [64] suggest that overfitting resistance of the AdaBoost algorithm is an artefact of using different criteria for training and testing the algorithm.

(d) Efforts on explaining the success of boosting algorithms in practical applications are their margin interpretation by [84], and their similarity to random forests as self-averaging interpolating algorithms [104].

(e) The similarity between AdaBoost and gradient boosting has been clarified in [64]. The latter study notes that the primary idea of AdaBoost is to up-weight misclassified observations. On the contrary, gradient boosting focuses on identifying and modelling large errors resulting from the previous iterations.

(f) Stochastic gradient boosting [36] is a development of gradient boosting. In each iteration of stochastic gradient boosting, a random subsample of the training data is used for training. This procedure increases robustness.

(g) When implementing gradient boosting algorithms with decision trees, the relative importance of predictor variables in predicting the response variable can be estimated [35]. Relative importance allows the ranking of the predictor variables with respect to their importance in predicting the response variable. Partial dependence plots, which show the effect of a variable in predicting the response variable after considering the average effects of the other predictor variables, are also possible [35].

## 2.3 Important developments of statistical boosting

During the evolution of boosting algorithms from machine learning to statistical modelling [64], numerous boosting extensions were suggested and new possibilities appeared for improving modelling procedures. Important developments include component-wise gradient boosting [17]. This type of boosting allows the modelling of each predictor variable using diverse base learners. For instance, one could model the first predictor variable $x_1$ using a decision tree, and the second predictor variable $x_2$ using a linear model. Later developments of this approach include procedures for high dimensional problems [13,14,43]. The generic framework was presented by [15].

The procedure of component-wise gradient boosting allows the modelling of different effects using different base learners, thereby resulting in interpretable models. Furthermore, the variable selection procedure is inherent to the framework, since at each iteration a single base learner (corresponding to a single predictor variable) is selected [63]. The framework can also be used for automatic model selection, in a similar way to that for variable selection [63]. Furthermore, using a single base learner for modelling blocks of predictor variables is also an option [63]. Related software implementing these developments can be found [42,44].

Generalized Additive Models (GAM) assume an exponential family distribution for the response variable, while the mean of the response variable is modelled as a function of the predictor variables. Likelihood-based boosting [95] was designed to circumvent problems appearing in generalized additive modelling (e.g. estimation in cases with a high number of predictor variables) by adopting stage-wise training of the base learners. This is done by maximizing an overall likelihood in each iteration [64].

### 2.4 Loss functions

Statistical boosting algorithms are flexible enough so that they can be optimized using diverse loss functions. The selection of a loss function depends on the problem that the researcher intends to solve. A non-exhaustive list of commonly used loss functions can be found in [42,67]. This list includes the loss functions $L_1$, $L_2$, the Huber loss function and the negative Gamma log-likelihood function with logarithmic link. These loss functions could be used when one aims to provide point predictions. The quantile and the expectile loss functions can be used when one aims to deliver probabilistic predictions. For categorical responses, one can use the binomial, the AdaBoost and the Area Under the receiver operating characteristic Curve (AUC) loss function. Loss functions for survival analysis (the negative partial log-likelihood for Cox models) and counts data (the Poisson loss) also exist.

### 2.5 Base learners

The selection of the base learner depends on the nature of the problem that one wishes to solve. A non-exhaustive list of base learners can be found in [42,67]. For instance, base learners can be linear models (ordinary linear regression, ridge penalized linear regression, random effects), smooth models (p-splines, radial basis functions), decision trees (tree stumps, decision trees in general), Markov random fields for spatial problems and wavelets, while relevant applications include time series models [86].

### 2.6 New techniques with high predictive performance

Recently, the interest in boosting algorithms from a predictive performance perspective has been revived. This revival has been made possible by the introduction of the Extreme Gradient boosting (XGBoost) algorithm [20]. This latter algorithm has been proved as a highly competitive one in machine learning competitions. The primary reason for its success is its scalability (i.e. its ability to handle cost-efficiently more and more workloads). Following the success of XGBoost, several new competitive algorithms that are based on similar ideas, were introduced. Such algorithms are Light Gradient Boosting Machine (LightGBM, [50]) Categorical Boosting (CatBoost, [74]) and Accelerated Gradient Boosting (AGB, [8]), among others.

## 3. Properties of boosting algorithms

Boosting is accompanied by several strengths stemming from its own nature, while it also retains the strengths of its base learners. For its better exploitation, it is important to know and understand these strengths, as well as the weaknesses of the algorithm. The properties of the algorithm are largely known from theoretical studies. They are also empirically understood through the extensive use of the algorithm during the last two decades. A summary of these properties can be found in the following.

### 3.1 Ease of use

1. Boosting is an "*off the shelf*" algorithm, i.e. it does not need extensive preprocessing and hyperparameter tuning to perform well [11,39 p.352].

2. Several of its variants have been implemented in open-source software; therefore, they are freely available. Existing software is flexible enough so that customized models can be implemented [63,64]. The modular nature of the boosting algorithms and their software implementations makes it even more appealing [15,42,44,62,65].

### 3.2 Generalization ability

3. Boosting algorithms are more accurate than single decision trees, and very accurate in general [35,62,64,67].

4. There are several regularization procedures to avoid overfitting, e.g. subsampling, shrinkage and early stopping (through cross-validation) [15,62,64,65,67], albeit boosting itself is also resistant to overfitting [15,17,64].

5. Boosting algorithms are robust against multi-collinearity issues [62].

6. They are memory consuming [67].

7. They are time-consuming, due to the number of iterations needed for proper training [35,62,65].

8. They can be time-consuming when training, since sequential fitting is required in principle; therefore, benefits from parallelization are not always possible [65,67].

9. They are time-consuming when they predict, because all base learners must be evaluated. That would be a problem e.g. in online tasks [67].

10. Recent advances have merely addressed issues of parallelization [64,65].

11. Boosting improvements in predictive performance are rapid at the initial iterations, while they level off to slower increments. Therefore, boosting can be used for a preliminary approach to the problem, aiming at understanding whether there is potential for further improvements, and later be implemented more intensively [35].

12. The first glance offered by boosting can be beneficial to understand the problem at hand and implement other algorithms later [35].

13. Boosting can be stopped, and the previous computations can be used when new data arrive, albeit this solution would be suboptimal compared to the case that the full dataset is used for training [35].

14. Recent advances of boosting in the machine learning community have managed to provide boosting algorithms (e.g. extreme gradient boosting, categorical boosting, light gradient boosting [20,50,74]) with unprecedented power and speed.

## 3.3 Flexibility

15. Because of its flexibility, boosting can be considered as a framework, thereby giving the opportunity to the user to define his/her model for a specific task [15,63,64,65,67].

16. Boosting algorithms can be non-parametric, semi-parametric or parametric (based on the choice of the base learner and the type of the dependent variable or parameter), depending on the specific requirements of the problem [15,62, 63,64,67,95].

17. Boosting algorithms can be combined with various base learners; thus, they can solve a variety of problems, ranging from typical machine learning problems (regression, classification) to less common ones (time series analysis, spatial analysis, density analysis, survival analysis, modelling of longitudinal data, functional regression) [15,17,42,44,62,63,64,65,67,86].

18. Diverse types of effects for each predictor variable can be modelled and estimated simultaneously (e.g. linear effect for one predictor variable and a decision tree for the second one) [42,44,62,63,64,67].

19. Likelihood-based boosting can be used to estimate parameters in a likelihood setting [63,64,95].

20. Boosting can be combined with different loss functions [15,17,42,44,62,63,64,65,67].

### 3.4 General properties

21. Boosting inherits the appealing properties of the base learners (usually decision trees), while it mitigates the negative ones [35].

22. It is robust with respect to the presence of outliers in the dependent variable when used with decision trees [35].

23. It can handle missing values when using decision trees as base learners [35].

24. It is invariant under strictly monotone transformations of the predictor variables when using decision trees as base learners [35].

25. It is robust, not being sensitive to outliers and heavy-tailed distributions of the predictor variables when using decision trees as base learners [35].

26. It can be regarded as an optimization tool [43,65].

### 3.5 Predictor variable selection

27. Boosting algorithms can generalize well in the presence of redundant predictor variables [17,35,64].

28. They can be used in high dimensional problems (in which predictor variables are far more than the observations). In these problems, a variable-selection procedure is automatically implemented (in case that additive base learners are used) [15,17,42,43,44,62,63,64,65,67,95].

29. In addition to their use in variable selection problems, they can also be used for model selection [15,62,63,64,65].

30. However, it may be preferable to use common models in low-dimensional problems [62].

### 3.6 Interpretability

31. They can be more interpretable compared to black-box machine learning models, due to their recent evolution resulting in statistical models [42,44,62,63,64,65].

32. Depending on the base-learners used in additive modelling (linear effects, splines and more), they can be interpretable. This holds since the effect of each variable can be

estimated independently [42,43,44,62,63,64,65,67].

33. Error estimates for resulting effects are not available in principle. Therefore, confidence intervals for effects cannot be estimated directly [62].

34. Boosting algorithms are less interpretable than single decision trees [35,67].

35. Interactions of the predictor variables can be modelled [43,67].

36. Relative importance of the predictor variables can be computed; therefore, boosting algorithms can be interpretable [35,67].

37. Boosting can be used for computing partial dependence plots, thereby exploiting some of its potential for interpretability [35,67].

## 4. Quantitative analysis of boosting applications in energy research

Here we analyse articles that include applications of boosting algorithms in energy research. To this end, we conducted a literature survey using the Scopus database (including papers appearing until 2020-02-29). In particular, we searched for papers meeting three conditions simultaneously. Specifically, the papers should be (a) published in the subject area of Energy (also including papers from the traditional Energy Journal, which is not assigned in this subject area), (b) published until year 2019, and (c) citing at least one of the following 11 papers (the number of citations is in parentheses): [15] (0), [20] (34), [31] (5), [32] (4), [34] (28), [35] (98), [37] (11), [50] (6), [74] (1), [82] (11), [95] (0). We consider these latter papers as key papers, as they are related either to the machine learning view of boosting [31,32,34,82], or the statistical view of boosting [15,35,37,95], or the new generation of algorithms with high predictive performance [20,50,74]. The original dataset (composed by papers meeting the three conditions above) includes 202 papers. The final dataset includes 157 papers (see the list of publications uploaded as supplementary material), i.e. those that remained after excluding (from the original dataset) papers that do not include boosting applications. We consider that our sample is representative of energy-related-boosting studies. Journals not belonging to the subject area of Energy of the Scopus database may occasionally publish papers related to Energy; however, we consider that including those Journals in our analysis would not result in significant improvements.

We thoroughly examined the 157 finally selected papers. In the remainder of Section 4, we present a quantitative analysis of our findings. Moreover, Section 5 complements Section 4 by presenting a qualitative analysis focusing on some special cases. The latter could not be revealed by our quantitative analysis.

## 4.1 General evaluation

To understand the main issues addressed when using boosting algorithms in energy research, we created the word cloud of Figure 1. Figure 1 is a visual representation of the frequency of words appearing in the titles and keywords of the 157 examined papers; thus, it can serve as a basis for the rest of our analysis. The word cloud has been created by grouping similar words under a single term. For instance, "prediction", "predictive" and "predicting" are represented by the term "prediction".

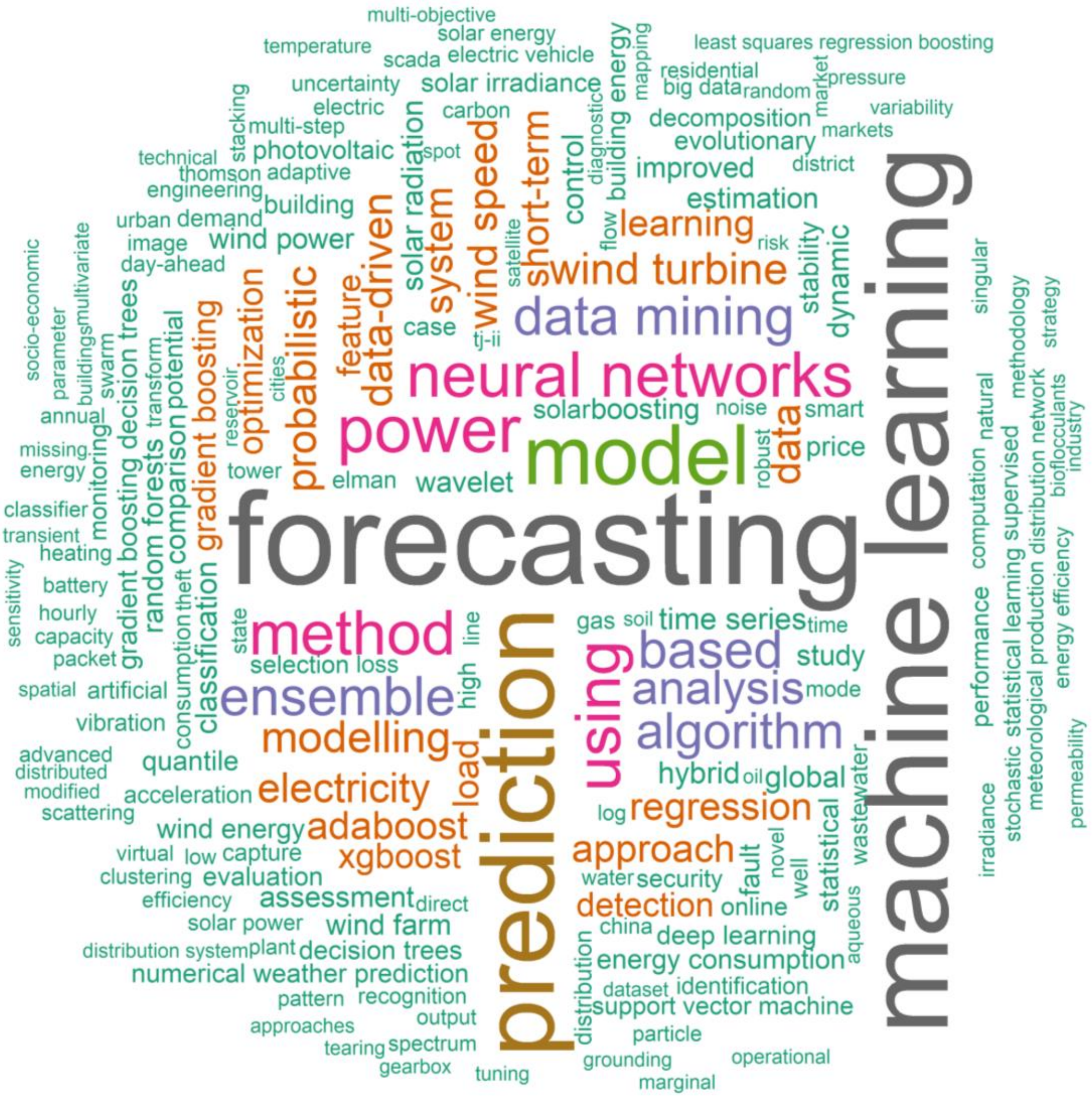

Figure 1. Word cloud based on titles and keywords of the 157 examined papers. Words that appear more frequently are given greater prominence (i.e. bigger font size).

One can conclude that the most frequently examined themes are related to "forecasting", "prediction", "regression", "analysis", "data mining" and "modelling" in general. Of particular importance is the frequent appearance of the term "probabilistic", which is mostly connected to forecasting (e.g. probabilistic forecasting). We should also emphasize that most studies focus on methods and applications (with the word “using” appearing frequently, while the terms “method” and “approach” are also frequently met). The frequent use of the term “machine learning” implies that the authors perceive boosting as a machine learning algorithm, rather than as a statistical one.

Terms related to the kind of the modelled energy type also appear frequently (e.g. wind speed, wind turbine, electricity, power). The kind of algorithms used is also emphasized with AdaBoost and XGBoost appearing frequently, while the term “neural networks” is also frequently met. The presence of the latter term may be related to the fact that such applications require some kind of benchmarking or comparison, in which neural networks seem to be the most frequently used benchmark model. The term “ensemble” also appears frequently, perhaps because boosting is an ensemble learner. Furthermore, its use as a combiner of diverse algorithms in a stacking framework is frequent as well, given that an important boosting property is optimization.

## 4.2 Specific issues

We conducted further investigations to examine specific issues and discover patterns associated with boosting applications in the dataset. Boosting applications remained constant per year of publication until year 2015 (see Figure 2). From year 2016, the number of applications increases exponentially. This might be related to the introduction of XGBoost and related high performance algorithms (i.e. LightGBM, CatBoost).

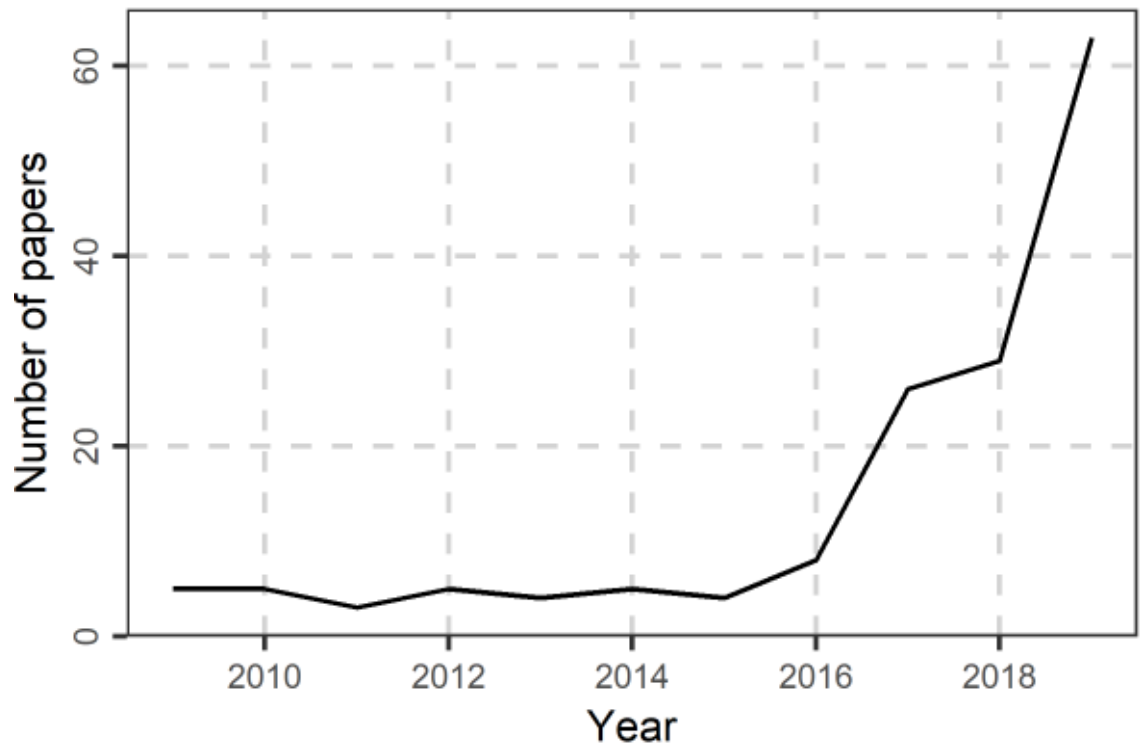

Figure 2. Total number of articles implementing boosting algorithms per year of publication.

Figure 3 presents the wide range of Journals that published the articles composing our dataset. Energies Journal includes more than 25 articles followed by Applied Energy Journal with 15 articles. We should consider that the number of boosting articles might be proportional to the number of articles published by the Journals. For instance, the Energies Journal publishes a high number of articles per year. The sample of Journals in the vertical axis of Figure 3 includes both Journals of general interest with respect to their energy type specialization (e.g. Energies, Applied Energy) and more specialized Journals (e.g. Journal of Petroleum Science and Engineering, Solar Energy).

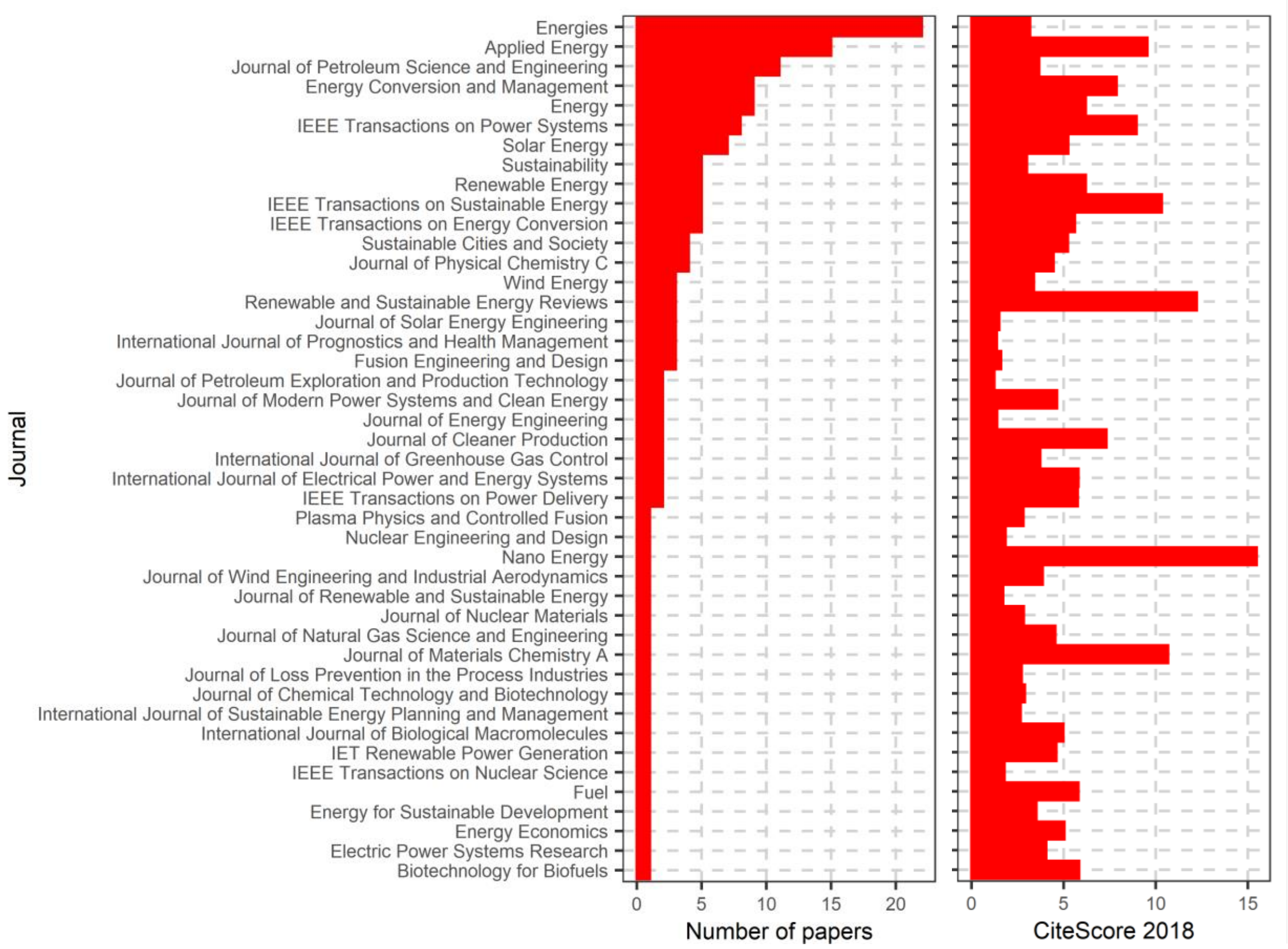

Figure 3. Number of published articles implementing boosting algorithms conditioned on the Journal (on the left) and Journal CiteScores (on the right).

Boosting algorithms have been applied to address a wide range of issues with respect to the examined energy type (see Figure 4). These issues mostly include wind energy and solar energy ones (see also discussion in Section 4.1). Electricity grids and electricity in general are also frequent themes, while almost every energy type is covered within our dataset. In the following visualizations (including Figure 4), it may happen that the sum of the examined variables of interest may be higher than 157. For instance, in Figure 4 some studies examine both wind energy and solar energy; therefore, they are counted in two energy types.

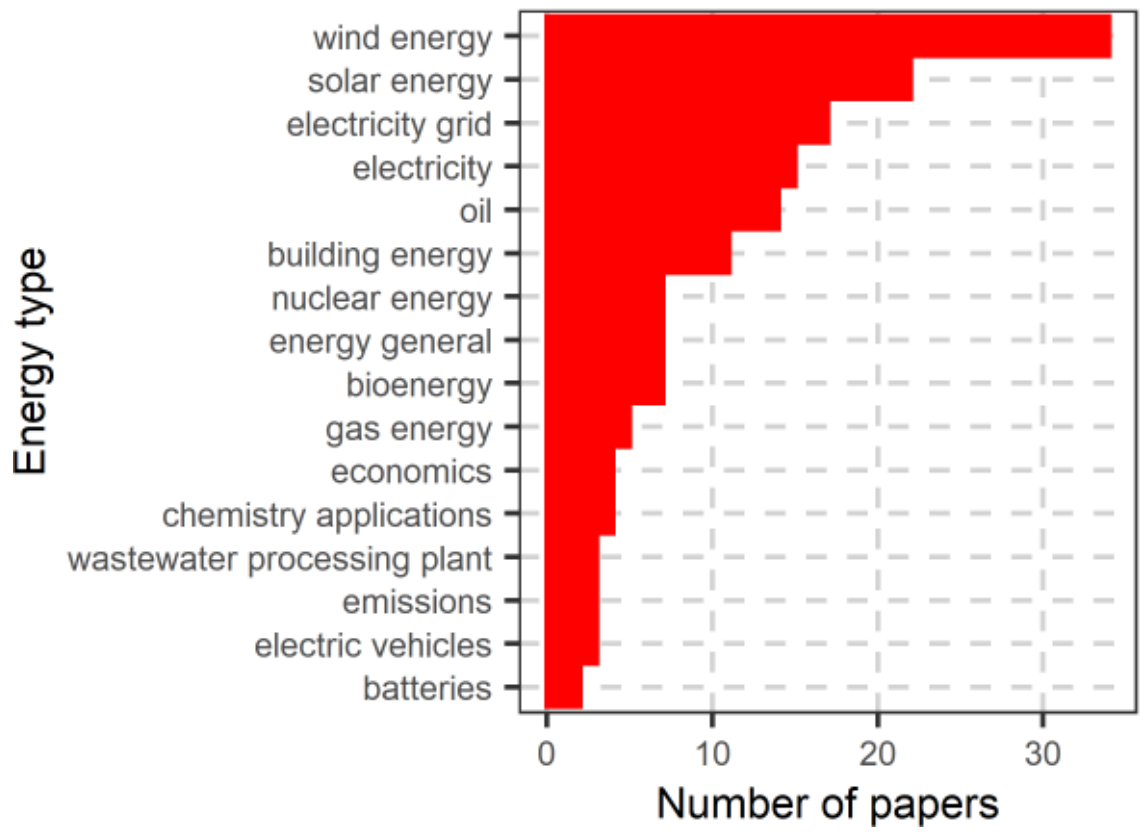

Figure 4. Number of published articles implementing boosting algorithms conditioned on the examined energy type.

Figure 5 presents the regions of origin for the data exploited in the examined articles. Articles based e.g. on laboratory experiments or simulation studies are grouped according to the origin of the authors. The majority of the published papers are associated with energy research conducted in/for the USA and China, while other regions outside Europe are mostly under-represented and Iran holds the second prominent position amongst the Asian countries. Some global datasets have been also exploited.

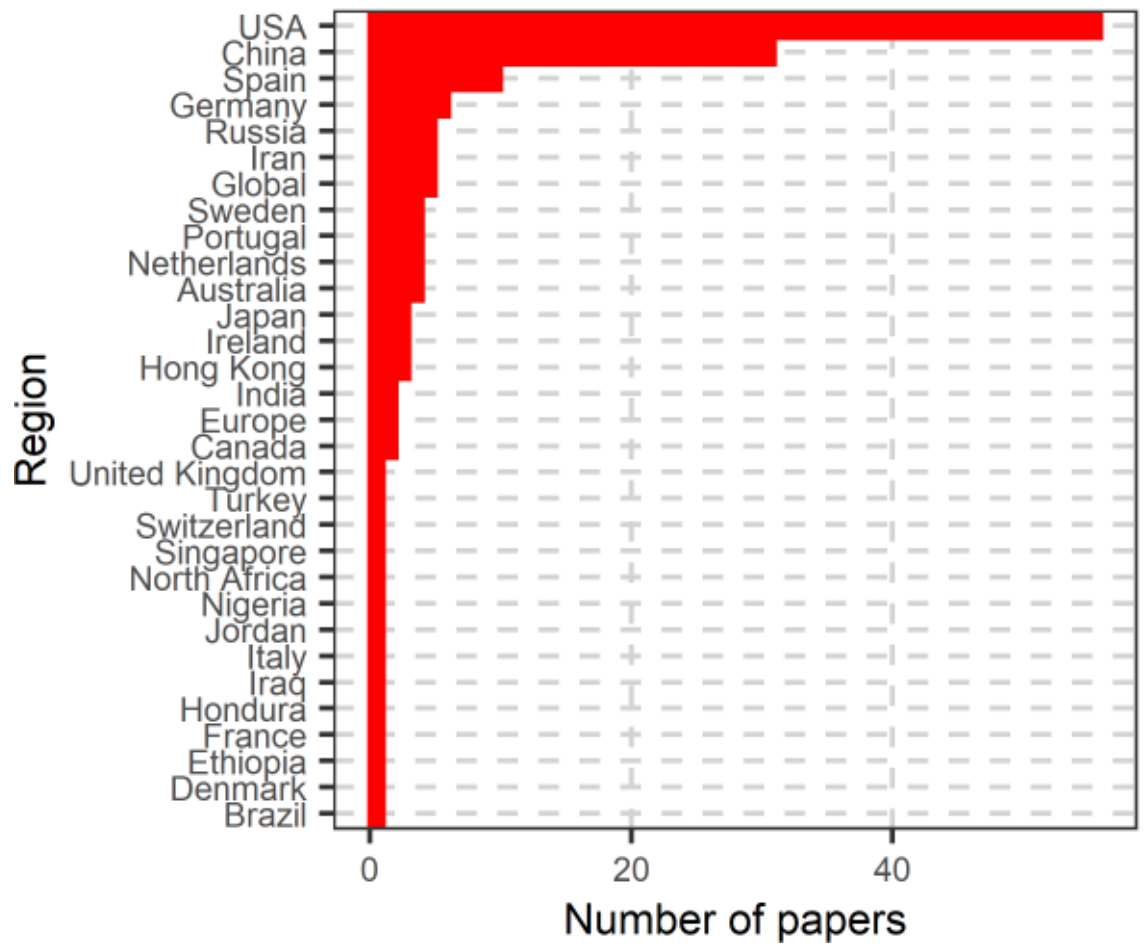

Figure 5. Number of published articles implementing boosting algorithms conditioned on the region of the application.

Most published papers use decision trees as base learners, followed by those using neural networks (Figure 6). This is not surprising, given the fact that large part of the boosting literature is associated with decision trees.

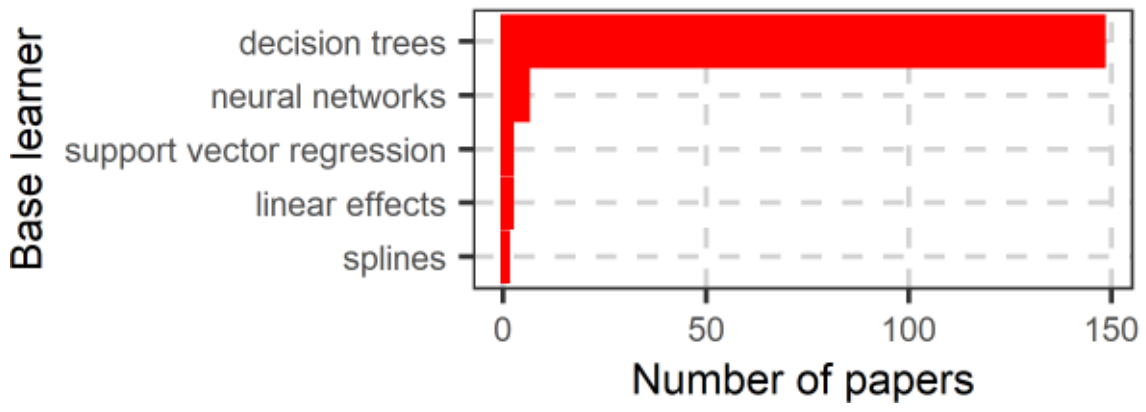

Figure 6. Number of published articles implementing boosting algorithms conditioned on the base learner.

Figure 7 presents the type of loss function used in the examined articles. Most articles use the loss function $L_2$, i.e. they are associated with regression tasks. Classification with the AdaBoost algorithm and classification with the binomial loss function (LogitBoost algorithm) are frequently implemented. The quantile loss related to the estimation of conditional quantiles (e.g. probabilistic forecasting) is also frequently met. We decided to group XGBoost, LightGBM and CatBoost applications separately, regardless of the implemented loss functions. This decision was taken because of their importance. XGBoost appears frequently, perhaps due to its hype gained by its high performance in machine learning competitions.

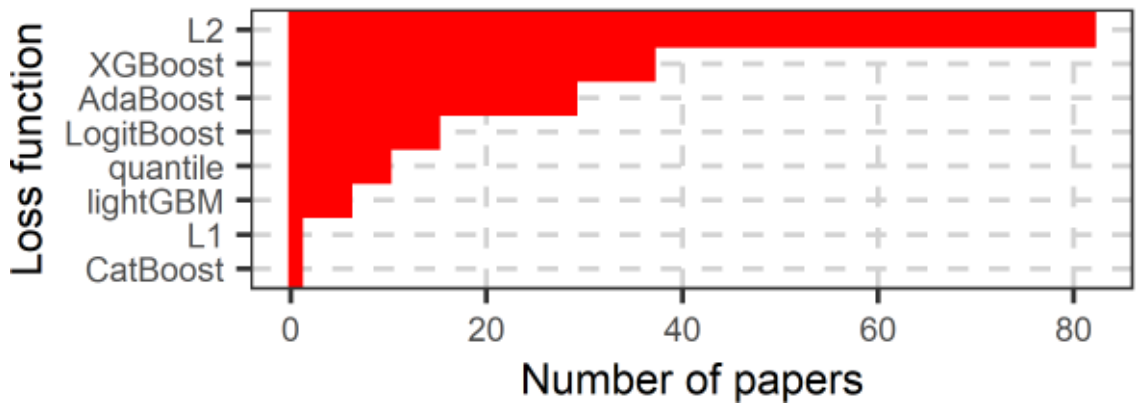

Figure 7. Number of published articles implementing boosting algorithms conditioned on the type of the loss function.

Figure 8 presents the types of applications appearing in the examined articles. Most applications are related with prediction and forecasting (both applications are essentially regressions). The term "prediction" refers to the case of regression in general, while we use the term "forecasting" for studies which aim to predict future values in a regression (with time series) framework. Classification is met less frequently in the examined studies. On the other hand, relative importance is frequently computed; however, it is surprising that related concepts from statistical view of boosting (e.g. automatic variable selection and more) are not met.

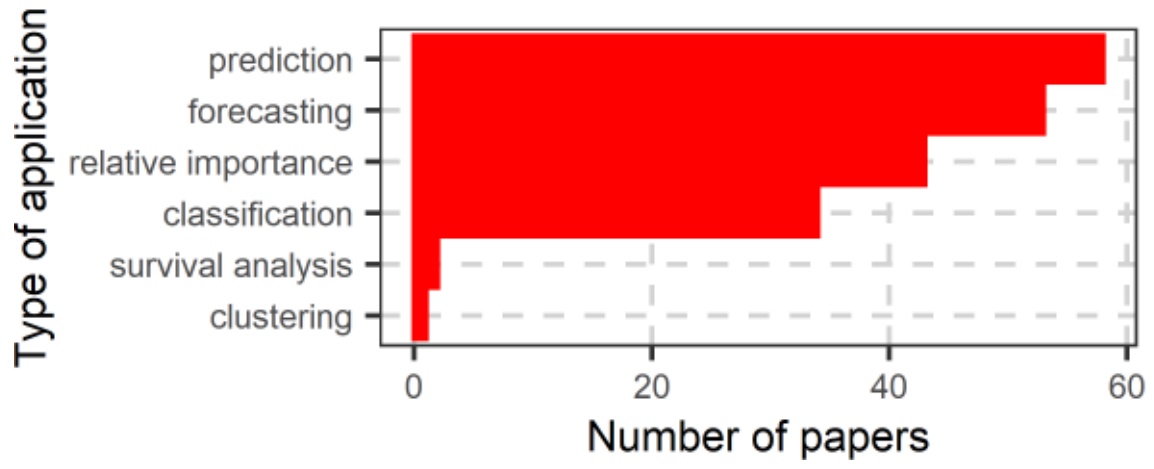

Figure 8. Number of published articles implementing boosting algorithms conditioned on the type of the application.

Studies conducted in a regression context comprise the vast majority of applications followed by those conducted in a classification context. Unsupervised learning (clustering) is rarely met (Figure 9).

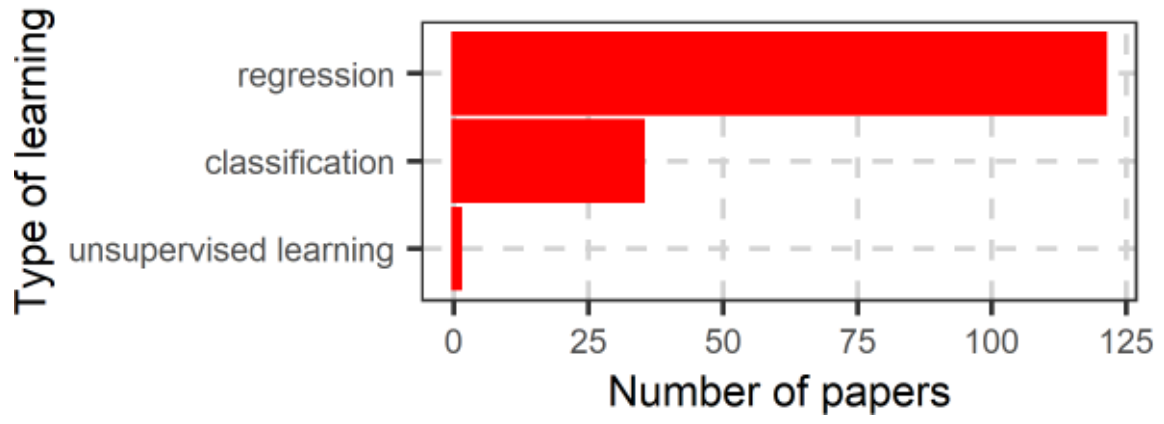

Figure 9. Number of published articles implementing boosting algorithms conditioned on the type of the learning task.

The allocation of papers to Journals based on the examined energy type is presented in Figure 10. Among these Journals, there exist some specialized in specific energy types (e.g. wind or solar energy, petroleum engineering). Multidisciplinary Journals (e.g. Energies Journal, Applied Energy Journal) are more balanced with respect to the examined energy type.

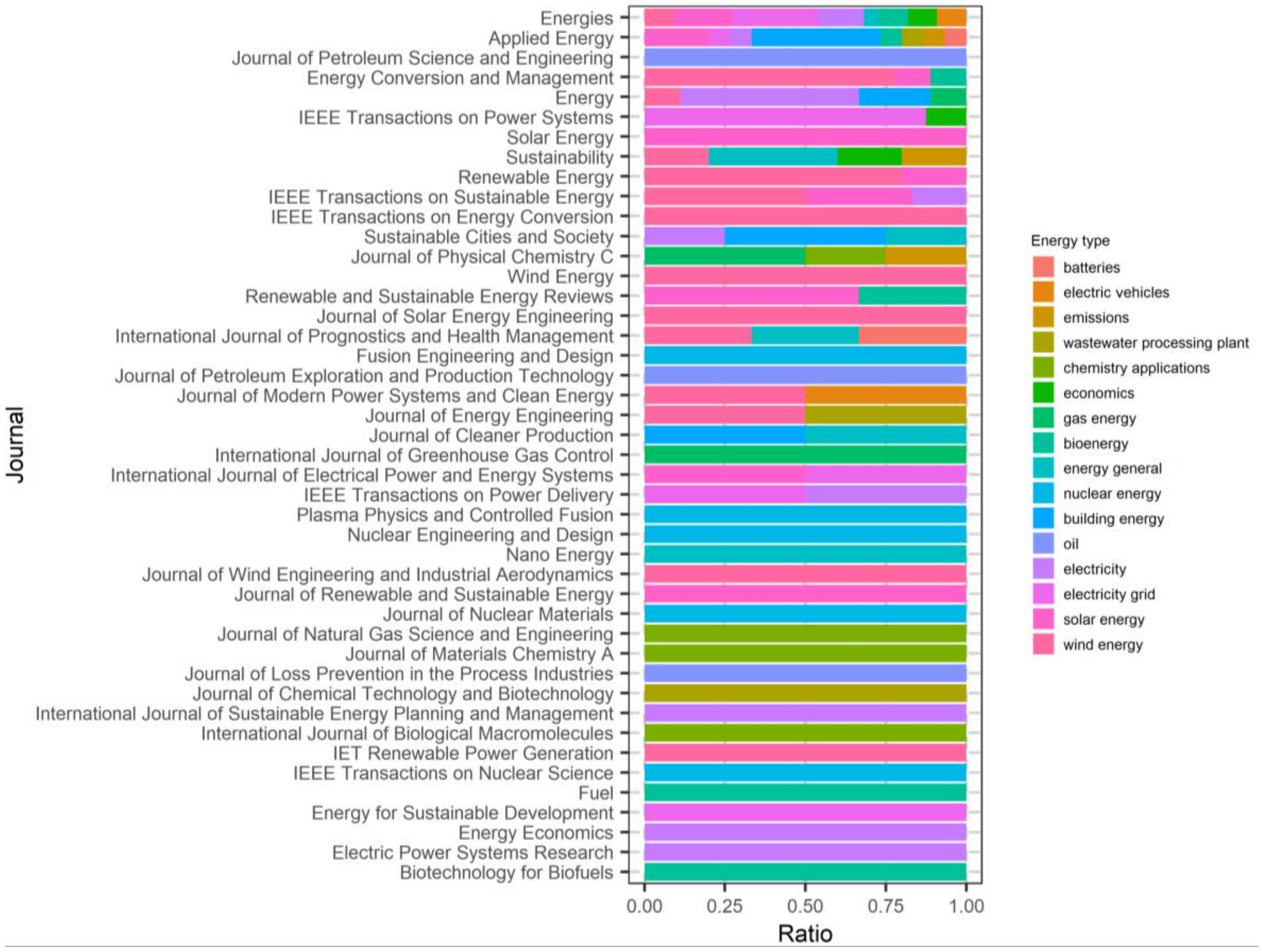

Figure 10. Ratio of the examined energy types conditioned on the Journal. The respective numbers of papers per Journal are shown in Figure 3.

The type of application conditioned on the specific energy type is presented in Figure 11. Studies on wind energy, solar energy and electricity focus on forecasting applications. Classification applications are mostly the focus of studies on electricity grids. Prediction applications are allocated evenly among studies examining the various energy types. Two applications of survival analysis can be found in studies examining nuclear energy.

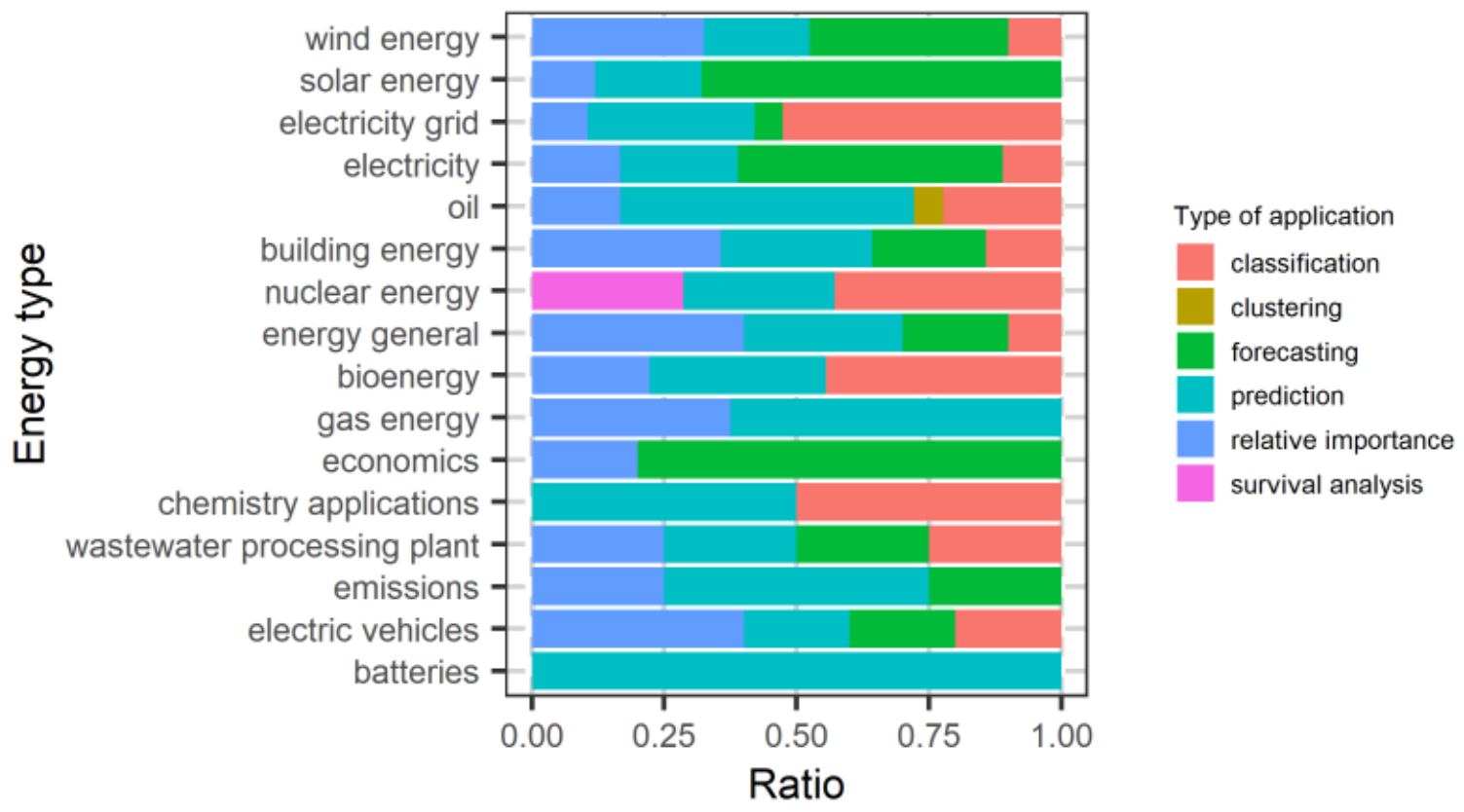

Figure 11. Ratio of the types of the examined applications conditioned on the examined energy type. The respective numbers of papers are shown in Figure 4.

## 5. Qualitative analysis of boosting applications in energy research

The primary use of boosting algorithms is made within statistical frameworks, mostly forecasting (e.g. electricity forecasting, wind speed forecasting, solar irradiance forecasting) frameworks. Such frameworks are trivial and straightforward to apply. They necessarily involve the selection of predictor variables, which range from lagged time series variables to seasonal effects. Furthermore, forecasting frameworks usually involve a comparison of some statistical and machine learning algorithms. Such comparisons are common independently of the algorithm of focus (e.g. neural networks, support vector regression). In the following, we will refrain from analysing such frameworks since energy researchers are familiar with them. Instead, we will focus on some alternative aspects of the boosting applications.

Based on the analysis of Section 4, we identified some special issues that need further consideration within the context of boosting applications in energy research. These themes are mostly related to the history of the first applications, to the type of the modelled energy variable, as well as to several algorithmic issues and special cases. In the following, some studies are analysed in more detail. In our opinion, these studies are representative for the examined special issues.

### 5.1 Early applications

The earliest applications of boosting algorithms in energy research trace back to 2009 (in spite of the fact that boosting was introduced much earlier). These applications focus on wind energy, with the group of Kusiak and co-authors being involved [52,53,54,55,112]. (Note that Kusiak's group is involved in 17 published papers related to boosting). These early studies have a strong engineering content, with boosting playing a secondary role, i.e. being used as an element of a larger framework. Therefore, they deviate from subsequent publications that have a strong statistical content. The latter category of studies involves e.g. the forecasting frameworks mentioned earlier in this Section. The first application in which boosting plays a primary role is again conducted by the group of Kusiak and co-authors [51]. In this latter application, boosting is used for forecasting wind speed in comparison to multiple statistical and machine learning models. Neural networks outperformed other methods in this application. The earliest boosting application that is not directly related to energy addresses the issue of wastewater plant energy consumption [109].

## 5.2 Electricity

We found some interesting electricity-related studies using boosting. An application to characterizing household features based on electricity demand data highlights the potential of boosting implementations for solving energy – socioeconomic issues [6]. Multidisciplinary issues have been addressed by boosting algorithms. This has been made by examining questionnaires in cross-sectional surveys. In such surveys, the number of predictor variables is high and the problems are more complex [22,26].

Another interesting issue addressed by boosting is related to grounding systems, and in particular grounding topology [89]. An important –yet rarely met– theme in machine learning applications is handling imbalanced datasets, i.e. cases where we aim to predict a class that is under-represented in the dataset. Such cases can be addressed by boosting algorithms, e.g. by assigning higher costs to the minority class. A related interesting application can be found in [49]. This latter study attempts to predict thunderstorm-induced power outages.

## 5.3 Wind and solar energy

The most frequent themes addressed by boosting algorithms are wind energy and solar energy themes. Such are mostly related to forecasting issues (earliest applications include [51] and [108], respectively); however, alternative applications are also met. Among them, we found interesting the classification application on bird recognition around a wind farm [106]. Another interesting application regards again an imbalanced dataset. In this latter application, random under-sampling combined with boosting was implemented for detection of non-technical losses in electricity grids [4]. An application on probabilistic solar power forecasting can be found in [7].

## 5.4 Other energy types

Besides electricity, wind energy and solar energy, other energy types have been examined using boosting, albeit to a lesser extent. An interesting study on probabilistic electricity price forecasting (economics of energy) can be found in [3]. Other studies include building energy consumption [78], survival analysis in nuclear energy [70,71], heat load (in which XGBoost is used as benchmark) [93], applications in chemistry [94] and applications in nanotechnology [102].

An interesting study in which boosting is combined with reinforcement learning to optimize performance in energy consumption in a wastewater treatment plant can be found in [29]. Boosting is used to substitute costly experiments in a study on irradiated metals [48]. Furthermore, some multidisciplinary studies on water-electricity demand nexus using multiple sources of data and a socio-economic study on carbon dioxide emissions can be found in [68] and [91], respectively.

### 5.5 Post-processing applications

An alternative to directly forecasting weather variables (wind speed, solar irradiance) by only using machine learning algorithms and historical information is to additionally exploit information from numerical weather predictions (which are relatively accurate in short-term horizons) within post-processing frameworks. Post-processing usually improves forecasts, while it can also transform point forecasts to probabilistic forecasts. Given its high predictive accuracy and its ability to provide probabilistic predictions, boosting is suitable for post-processing applications. Such applications include [2] and [56], among others. The former application example combines numerical weather predictions and provides probabilistic forecasts, and the latter application example uses satellite data and post-processing of numerical weather forecasts for short-term forecasting of solar irradiance. The works by [5] and [10] are also of interest. These works compare multiple methods, including boosting, for probabilistic post-processing of solar radiation and energy consumption / production, respectively.

### 5.6 Stacking frameworks

Boosting can be regarded as an optimization tool; therefore, it can be effectively used in stacking frameworks by combining several algorithms and by optimizing their results with respect to a loss-function. Such studies include combinations based on AdaBoost for wind speed prediction [59], boosting for short-term electricity consumption forecasting [25], and XGBoost for gearbox fault prediction of wind turbines [107] and short-term photovoltaic power output prediction [114].

### 5.7 Large-scale studies and big datasets

Of particular importance are large-scale studies using multiple algorithms, studies using big datasets and forecasting competitions. Such empirical studies can provide understanding of the properties of the involved algorithms. They are rarely met in energy research, perhaps due to limited data availability. In [1], XGBoost is used for combining multiple physics-based models for solar irradiance modelling in 54 stations located worldwide. Multiple boosting algorithms are compared in forecasting electricity demand using multiple datasets [110]. Multiple boosting methods are also used in a building energy application [24]. Moreover, multiple algorithms are compared for long-term global solar radiation prediction in [38]. The largest study with respect to number of its compared algorithms is [105]. This latter study compares 68 methods in hourly solar forecasting. Finally, global datasets have been handled by boosting algorithms in bioenergy related studies [46,47].

### 5.8 Interpretability

Most of the examined studies focus on issues related to predictive performance, and only a few ones focus on interpretability and attempt to explain observed behaviours. The work by [28] applies model-agnostic methods to explain data-driven building energy performance models, including XGBoost. However, the primary use of boosting algorithms is computation of relative importance (see e.g. the XGBoost application in [92]). An application involving many predictor variables can be found in [103]. This latter work focuses on understanding the properties of metal – organic frameworks.

### 5.9 XGBoost and related algorithms

An extended comparison of XGBoost with other models (including physics-based) in global solar radiation modelling can be found in [98]. XGBoost has also been combined with a time series model (to form a hybrid model) for energy supply security forecasting [57]. Finally, one study implements XGBoost, LightGBM and CatBoost to model non-technical loss detection of an energy system [75].

### 5.10 Other uses of boosting algorithms

Model-based boosting has been used as a benchmark model, i.e. new models were compared against it for probabilistic electricity load forecasting [58]. Furthermore, the work by [61] applied boosting to classify counties in US states according to their involvement in green buildings projects. Finally, a new algorithm for classification, based on AdaBoost and using neural networks as base learners, has been proposed by [76] for nuclear energy systems.

## 6. Concluding remarks

We conclude with the following remarks and issues for future consideration:

a. Since their introduction in the early 90s, boosting algorithms have progressed. The principal progress has been observed in the field of statistical boosting, thereby moving from the machine learning mind-set towards statistical thinking. On the other hand, boosting algorithms oriented to issuing state-of-the-art predictions have appeared.

b. The machine learning view of boosting mostly focuses on properties related to its high predictive performance and its good generalization ability. New boosting algorithms (XGBoost, LightGBM, CatBoost) excel at such tasks.

c. The statistical view of boosting focuses on properties related to its interpretability, and its potential to be applied by using a diverse set of base learners, and variable and model selection.

d. Boosting algorithms have some weaknesses, e.g. they can be time- and memory-consuming. Recent developments try to address these issues.

e. Papers using boosting algorithms in energy research are devoted to a wide range of themes, including renewable energy, electricity, fossil fuels, nuclear energy and bioenergy themes, among others.

f. Papers in energy research are mostly oriented in providing accurate predictions and forecasts. The vast majority of the examined papers are related to some type of data-driven forecasting. Such applications have become trivial, since they require a procedure of selecting predictor variables and then algorithmic training. Such procedures have been automated in recent software developments.

g. Given the primary focus in energy research, adoption of the new boosting algorithms (XGBoost, LightGBM, CatBoost) has been speeded.

h. On the other hand, applications with statistical orientation and focus on interpretation (as defined by the statistical view of boosting) have remained limited.

i. The latter conclusion should be considered along with the fact that most of the applications use relative importance metrics associated to boosting algorithms with decision trees as base learners for interpreting fitted models. While relative importance metrics are well known in the machine learning community (e.g. similar metrics exist for random forests), interpretability-related properties originating from statistical boosting developments seem to be neglected.

j. Energy research is limited to the use of single time series, while it is a common practice to examine a few cases. Perhaps limited data availability prohibits larger applications (albeit a few exceptions have been noticed).

k. Furthermore, data in multiple dimensions (e.g. spatiotemporal data) are even more limited. Consequently, boosting applications are limited to training a simple model, while the progress in combining multiple types of models (for instance time series and spatial models) has not been exploited.

l. In addition, high dimensional datasets (e.g. datasets with numerous variables) are also limited in energy, thereby further limiting the implementation of boosting models in energy research.

m. We anticipate that the increasing availability of diverse big datasets with multiple dimensions and a large number of features, combined with multidisciplinary approaches, will redefine the scope of boosting applications in energy, from a simple predictive algorithm to a more advanced statistical framework.

n. Some exceptions from the observed general patterns can provide some new pathways for fully exploiting the properties of boosting algorithms.

o. Such exceptions are, e.g. post-processing applications in which boosting algorithms are combined with physics-based models to improve prediction accuracy.

p. Since boosting can be regarded as an optimization tool, stacking frameworks (in which boosting is used to combine multiple models) have also been observed.

q. Boosting applications using big datasets and multidisciplinary boosting applications (using datasets obtained from multiple sources) have also been conducted and can serve as a guide for future research.

**Conflicts of interest:** We declare no conflict of interest.